\documentclass[floats,aps]{revtex4}
\usepackage{calc}
\usepackage{graphicx}
\begin{document}
\begin{abstract}
We calculate the dispersion of the out-of-phase mode
characteristic for the bilayer $\nu = 1$ quantum Hall system
applying the version of Chern-Simons theory of Murthy and Shankar
that cures the unwanted bare electron mass dependence in the
low-energy description of quantum Hall systems. The obtained value
for the mode
 when $d$, distance between the layers, is zero is in a good agreement with
 the existing pseudospin picture of the system. For $d$ nonzero but small we find that
the mode is linearly dispersing and its velocity to a good approximation depends linearly
on $d$.
This is in   agreement with the Hartree-Fock calculations of the
pseudospin picture that predicts a linear dependance on $d$, and
contrary to the naive Hartree predictions with dependence on the
square-root of $d$.
We set up a formalism that enables one to consider fluctuations around the found
stationary point values. In addition we address the case of imbalanced layers
 in the Murthy-Shankar formalism.
\end{abstract}
\title{Composite bosons in bilayer $\nu = 1$ system:
An application of the Murthy-Shankar formalism}
\author{Ivan Stani\'{c} and Milica V. Milovanovi\'{c}}
\address{Institute of Physics, P. O. Box 68, 11080 Beograd, Serbia and Montenegro}
\date{\today}
\maketitle
\vskip2pc
\narrowtext

A major problem surfaced in the early Chern-Simons (CS) composite
boson description \cite{ezi} of $\nu =1$ quantum bilayer. Namely,
a bare electron mass appeared in the Bogoliubov out-of-phase
gapless mode dispersion, which is unwanted due to the expectation
that the leading description of any intra-Landau-level collective
mode depends solely on interactions. The same requirement applies
for the description of any quasiparticles that may exist in the
lowest Landau level, namely their mass should stem from
interactions \cite{nre}. Murthy and Shankar \cite{sha,msh} put
forward an extended CS theory that was able to prescribe such a
mass for composite particles. The theory underwent major additions
and improvements (including an extension to higher momenta)
\cite{rev}, but in this paper we will use the early version
\cite{msh} best suited for our needs, i.e., the calculation of the
dispersion relation of the out-of-phase mode in a small-momentum,
low-energy window.

On the other hand, what is believed to be the first calculation of
the dispersion of the gapless mode in the scope of the pseudospin
picture (in which an electron can be in states that are
superpositions of localized layer states) was done by Fertig
\cite{fer}. He obtained an interaction-dependent velocity of a
linearly dispersing mode when $d$ is nonzero and a spin wave
quadratic dispersion with interaction-dependent spin stiffness
when $d$ is zero. The problem at $d = 0$ was again addressed in
Ref. \cite{son}
where a quantum ferromagnet picture for this case was established.
Then followed the pseudospin picture of reference Ref. 
\cite{moo},
a standard reference
 for the bilayer $\nu =1$ problem, in which small-momentum dispersion of the gapless mode was in
 agreement with Fertig's.

Our Hamiltonian is
\begin{equation}
{\cal H} = \sum_{i,\sigma} \frac{|\vec{p}_{i,\sigma} + e \vec{A}(\vec{r}_{i,\sigma})|^{2}}{2 m} +
V_{E} + V_{A},
\end{equation}
where $ \vec{A}$ is the vector potential of the constant external
magnetic field $- B_{o} \vec{e}_{z}$, so that the average total
density is $ n = \nu/2 \pi l_{B}^{2}$ with $ \nu = 1/(2 s + 1)$, $
s = 0,1, \ldots$ (for generality) and $ l_{B} = 1/\sqrt{e B_{0}}$,
the magnetic length. We take $ l_{B} = 1$ and $ \hbar  = 1$. $m$
is the ``bare'' electron mass, which is precisely the effective
mass  of electron in GaAs. $V_{A}$ and $V_{E}$ denote intralayer
and interlayer interactions, respectively. $ \sigma = \uparrow,
\downarrow$ is the layer index.

What follows is a brief account of a simple generalization of the
Murthy and Shankar CS theory extended to the case of the bilayer.
Missing details and explanations that are relevant also to the
single-layer case can be found in Refs. \cite{sha,msh,rev}.
As usual in the CS theory we make the unitary transformation
\cite{zha},
\begin{equation}
U = \exp\left\{i (2 s + 1) \sum_{i<j} \Phi_{ij}\right\},
\end{equation}
where $ \Phi_{ij} $ is the phase of the difference, $ z_{i} -
z_{j}$, of any two coordinates $z_{i}$ and $z_{j}$. Therefore, we
emphasize the transformation is the same, irrespective of the
layer indices. The corresponding Hamiltonian is
\begin{equation}
{\cal H}_{CS} = U^{-1} {\cal H} U = \sum_{i,\sigma}
\frac{|\vec{p}_{i,\sigma}+e \vec{A}(\vec{r}_{i,\sigma}) +
\vec{a}_{CS}|^{2}}{2 m} + V_{E} + V_{A} \label{csham}
\end{equation}
where the new ``gauge'' field satisfies the following connection
with the total density, $\rho(r)$:
\begin{equation}
\vec{\nabla} \times \vec{a}_{CS} = 2 \pi (2 s + 1) \rho(r)
\vec{e}_{z}. \label{cseq}
\end{equation}
Then we consider averaged and fluctuating values of $
\vec{a}_{CS}$ and $\rho$ \cite{zha}, rewriting Eq. (\ref{cseq}) as
\begin{equation}
\vec{\nabla} \times \langle\vec{a}_{CS}\rangle + \vec{\nabla}
\times : \vec{a}_{CS}: = 2 \pi (2 s + 1) n + 2 \pi (2 s + 1) :
\rho :
\end{equation}
so that equivalently, due to the cancelation of the external and averaged CS field, we can rewrite
 Eq. (\ref{csham}) as
\begin{equation}
{\cal H}_{CS} = \sum_{i,\sigma} \frac{|\vec{p}_{i,\sigma} + :
\vec{a}_{CS} :|^{2}}{2 m} + V_{E} + V_{A}.
\end{equation}

Now Shankar and Murthy \cite{sha,msh}, analogously to what Bohm
and Pines \cite{boh} did in the case of a three-dimensional (3D)
Coulomb gas, introduce magnetoplasmon degrees of freedom as
separate and elementary but necessarily satisfying certain
constraints with particle degrees of freedom in order to avoid
overcounting. They do this by introducing a pair of conjugate
fields, $ a(\vec{q})$ and $P(\vec{q})$, for each $\vec{q}$ in a
disk in the momentum space,
\begin{equation}
[a(\vec{q}),P(\vec{q}\ ')] = i (2 \pi)^{2} \delta(\vec{q} +
\vec{q}\ ').
\end{equation}
[reminding us (for fixed $\vec{q}$) of the harmonic oscillator
commutation relation in $(x,p)$ representation] and further,
defining a longitudinal and a transverse field, $\vec{P}(\vec{q})$
and $\vec{a}(\vec{q})$, respectively, as
\begin{equation}
\vec{P}(\vec{q}) = i \hat{q} P(\vec{q}) \; \; \; {\rm and} \; \; \;\vec{a}(\vec{q}) = - i \vec{e}_{z} \times
 \hat{q} a(\vec{q}).
\end{equation}
They rewrite the Hamiltonian (density) as
\begin{equation}
{\cal H} = \frac{1}{2 m} \sum_{\sigma} \Psi^{\dagger}_{CS,\sigma}
(-i \vec{\nabla} + :\vec{a}_{CS}: + \vec{a})^{2} \Psi_{CS,\sigma}
+ \tilde{V}_{A} + \tilde{V}_{E},
\end{equation}
in the second-quantized language, with the requirement
(constraint) that
\begin{equation}
a(\vec{q}) |{\rm physical} \; {\rm state}\rangle = 0,
\end{equation}
for each $\vec{q}$ such that $|\vec{q}| < Q$ where $Q$ is the
radius of the disk. Now it is convenient to eliminate :\
$\vec{a}_{CS}$: in favor of $a$ and $P$, and Shankar and Murthy do
that \cite{sha} applying the following unitary transformation,
\begin{equation}
U = \exp\left\{i \sum_{|\vec{q}|< Q} P(-\vec{q}) \frac{2 \pi (2 s
+ 1)}{q} \rho(\vec{q})\right\}
\end{equation}
where $\rho(\vec{q}) = \rho_{\uparrow}(\vec{q}) + \rho_{\downarrow}(\vec{q})$ i.e. the total charge
density. Now for the Hamiltonian density we have
\begin{equation}
{\cal H} = \sum_{\sigma} \frac{1}{2 m} \Psi^{\dagger}_{kb,\sigma} (- i \vec{\nabla} + \vec{a} +
2 \pi (2 s + 1) \vec{P} + \delta \vec{a})^{2} \Psi_{kb,\sigma} + \tilde{V}_{A} + \tilde{V}_{E}
\end{equation}
where $\Psi_{kb,\sigma}$'s denote the transformed fields, which describe modified, transformed
quasiparticles, and $\delta \vec{a}$ is the remnant of the CS field left uncanceled for
$|\vec{q}| > Q$.
 The constraint gets the following form:
\begin{equation}
\left[a(\vec{q}) - \frac{2 \pi (2 s + 1)}{q}
\rho(\vec{q})\right]|{\rm physical}\;{\rm state}\rangle = 0 \;\;
{\rm for} \;\; |\vec{q}| < Q.
\end{equation}
Neglecting \cite{sha} $\delta \vec{a}$
from the start and introducing \cite{sha}
\begin{equation}
A(\vec{q}) = \frac{a(\vec{q}) + i 2 \pi (2 s + 1)
P(\vec{q})}{\sqrt{4 \pi (2 s + 1)}}
\end{equation}
and
\begin{equation}
c(\vec{q}) = \hat{q}_{-} \sum_{j,\sigma} \vec{p}_{j,\sigma +}
\exp\{- i \vec{q}\cdot \vec{r}_{j,\sigma}\}, \label{cdef}
\end{equation}
where $V_{\pm} = V_{x} \pm i V_{y}$ for an arbitrary vector $\vec{V} = V_{x} \hat{e}_{x} + V_{y}
\hat{e}_{y}$,
we can rewrite our Hamiltonian as
\begin{equation}
{\cal H} = \sum_{i,\sigma} \frac{\vec{p}^{2}_{i,\sigma}}{2 m} + \sum_{|\vec{q}|<Q} \omega_{c}
A^{\dagger}(\vec{q}) A(\vec{q}) + \sqrt{\frac{\pi (2 s + 1)}{m}}
\sum_{|\vec{q}|<Q} [ c(\vec{q}) A^{\dagger}(\vec{q}) + c^{\dagger}(\vec{q}) A(\vec{q})] + V_{A} + V_{E}.
\end{equation}
As expected $\omega_{c} = e B_{0}/m$, i.e., equal to the cyclotron
frequency and in deriving the magnetoplasmon term we neglected
also total density fluctuations \cite{msh}.

Our quasiparticles are bosons and again for the sake of
completeness and easy reference we give brief account of the
so-called final representation in the Murthy-Shankar approach
applied to the case of two species of composite bosons. First
Murthy and Shankar always approximate as
\begin{equation}
\sum_{i,\sigma} \exp\{i (\vec{q} - \vec{k})\cdot \vec{r}_{i}\}
\approx n \; (2 \pi)^{2} \delta^{2}(\vec{q}-\vec{k})
\end{equation}
in the long-wavelength approximation so that also in this bosonic
representation we consistently have
\begin{equation}
[c(\vec{q}),c^{\dagger}(\vec{q}\ ')] \approx 0.
\end{equation}
To decouple the oscillators and particles they apply the following canonical transformation,
\begin{equation}
U(\lambda_{0}) = \exp\{i S_{0} \lambda_{0}\} = \exp\{\lambda_{0}
\theta \sum_{|\vec{q}|<Q} [c^{\dagger}(\vec{q}) A(\vec{q}) -
A^{\dagger}(\vec{q}) c(\vec{q})] \}, \label{fintra}
\end{equation}
where
\begin{equation}
\theta = \frac{1}{\sqrt{4 \pi (2 s + 1)} n},
\end{equation}
and the parameter $\lambda_{0}$ should be determined. As we have
new variables $\Omega$ defined through
\begin{equation}
\Omega^{old} = \exp\{-i S_{0} \lambda_{0}\} \Omega \exp\{i S_{0}
\lambda_{0}\}, \label{first}
\end{equation}
Murthy and Shankar also define
\begin{equation}
\Omega(\lambda) = \exp\{-i S_{0} \lambda\} \Omega \exp\{i S_{0}
\lambda\}, \label{second}
\end{equation}
so that in this case we have
\begin{eqnarray}
& A(\vec{q},\lambda) = A(\vec{q}) - \theta \lambda c(\vec{q}), & \nonumber \\
& c(\vec{q},\lambda) = c(\vec{q}), &  \;\;\; |\vec{q}| < Q.
\end{eqnarray}
It is easy to check that $\lambda_{0} = 1$ does the job of
decoupling and now we concentrate how variables $
\rho_{\sigma}^{old}(\vec{q}) $, $\sigma = \uparrow, \downarrow$
are connected with new ones. We use their definitions $
\rho_{\sigma}^{old}(\vec{q}) = \sum_{i} \exp\{ i \vec{q}\cdot
\vec{r}_{i,\sigma}\}$, $\sigma = \uparrow, \downarrow$ apply Eqs.
(\ref{first}) and (\ref{second}), find $ d
\rho_{\sigma}(\vec{q},\lambda)/d \lambda $, and integrate over
$\lambda$ to get
\begin{equation}
\rho_{\sigma}^{old}(\vec{q}) = \rho_{\sigma}(\vec{q}) +
\frac{q}{\sqrt{4 \pi (2 s + 1)}} \frac{n_{\sigma}}{n} \left[ [
A(\vec{q}) + A^{\dagger}(- \vec{q})] -
 \frac{\theta}{2} [c(\vec{q}) + c^{\dagger}(-\vec{q})]\right].
\label{sden}
\end{equation}
Immediately we can conclude that the spin density $ \rho_{s}^{old}(\vec{q}) =
\rho_{\uparrow}^{old}(\vec{q}) - \rho_{\downarrow}^{old}(\vec{q})$ is invariant,
\begin{equation}
\rho_{s}^{old}(\vec{q}) = \rho_{s}(\vec{q}),
\label{newsd}
\end{equation}
under the transformation of the final representation
[Eq. (\ref{fintra})], when we assume that we have the same fixed
number of particles, $n_{\uparrow} = n_{\downarrow} = n/2$, in
each layer. This is our main claim in the Murthy-Shankar formalism
for the bilayer system.

The analysis for the charge density and the form of the constraint
in new variables proceeds as in Refs. \cite{msh,sha}
and finally for the form we get
\begin{equation}
\rho(\vec{q}) = - \frac{i}{2} \sum_{j,\sigma} (\vec{q} \times
\vec{p}_{j,\sigma}) \exp\{- i \vec{q}\cdot \vec{r}_{j,\sigma}\}.
\label{newd}
\end{equation}
In a few lines but also using an assumption that we deal with an infinite system (without boundary)
we can prove that, in the second-quantized language, the constraint is
\begin{equation}
\int d\vec{r} \exp\{-i \vec{q}\cdot \vec{r}\} \rho(\vec{r}) =
\frac{i}{2} \int d\vec{r} \exp\{-i \vec{q}\cdot \vec{r}\}
\sum_{\sigma} [\vec{\nabla} \times \Psi^{\dagger}_{\sigma}
(\vec{r}) \vec{\nabla} \Psi_{\sigma}(\vec{r})] \label{voreq}
\end{equation}
for $|\vec{q}| < Q$ and as a shorthand notation we use
$\Psi_{kb,\sigma} \equiv \Psi_{\sigma}$ also in the following. The
proof starts by expressing the single-particle operator,
\begin{equation}
\sum_{j,\sigma} ( \vec{q} \times \vec{\nabla}_{j}) \exp\{- i
\vec{q}\cdot \vec{r}_{j}\},
\end{equation}
in the second-quantized language as
\begin{equation}
\sum_{\sigma}\int d \vec{r} \; \Psi^{\dagger}_{\sigma}(\vec{r})
(\vec{q} \times \vec{\nabla}) \exp\{-i \vec{q}\cdot \vec{r}\}
\Psi_{\sigma}(\vec{r}),
\end{equation}
then followed by simple regroupings and the neglect of a surface
term. So we find that in the long-wavelength approximation we use,
the charge density fluctuations exist only if there are vortex
excitations in the system. [The charge density on the right-hand
 side (rhs) of Eq. (\ref{voreq}) is propotional to the vortex
densities of the two kinds of fields. Strictly speaking the vortex
density is defined only by the phase part of a bosonic field, but
in the small-momentum limit we will neglect the difference. Later
a relaxation of the constraint and thereby generation of terms
quadratic in momenta will be justified by this difference.]
 If our system is a stable 2D Bose system that would mean that we have the case
for incompressibility, because to excite a vortex a finite energy is needed so all charge
fluctuations are supressed. But do we have a stable system? We have to get back to the Hamiltonian
expressed in new variables.
The Hamiltonian is
\begin{equation}
{\cal H} = \sum_{i=1,\sigma} \frac{\vec{p}_{i}^{2}}{2 m} -
\frac{1}{2 m n} \sum_{i,\sigma,j,\sigma^{'}} \sum_{|\vec{q}| < Q}
\vec{p}_{i,\sigma -} \exp\{- i \vec{q}\cdot (\vec{r}_{i} -
\vec{r}_{j})\}
                     \vec{p}_{j,\sigma^{'} +}
+ \omega_{c} \sum_{|\vec{q}| < Q} A^{\dagger}(\vec{q}) A(\vec{q}) + V_{E} + V_{A}.
\label{canh}
\end{equation}
To eliminate the bare electron mass in the kinetic energy and low
energy description we choose, as in Refs. \cite{sha,msh},
that the number of the oscillators is the same as the number of
particles, so that
 the diagonal part ($i = j$, $\sigma = \sigma\ '$) of the second term in Eq. (\ref{canh}) exactly cancels
the first kinetic energy term. The bare mass is still present in
the off-diagonal part of the second term, and if we decompose the
$q$ sum as \cite{sha,msh}
\begin{equation}
\sum_{|\vec{q}| < Q} \exp\{- i \vec{q}\cdot (\vec{r}_{i} -
\vec{r}_{j})\} = \delta^{2}(\vec{r}_{i} - \vec{r}_{j}) -
\sum_{|\vec{q}| > Q} \exp\{- i \vec{q}\cdot (\vec{r}_{i} -
\vec{r}_{j})\},
\end{equation}
we are left with a $\delta$-function interaction among particles
and another short-range interaction that may be grouped with
\cite{msh} previously neglected short-range pieces. We are
assuming all along that the same kind (layer) of bosons
(transformed electrons) behave as hard core bosons, so for $\sigma
= \sigma^{'}$ we see that the $\delta$ function is ineffective. To
eliminate the bare mass in the $\delta$-function interaction
between the opposite kind bosons we require that they also behave
mutually as hard core bosons. As we will see, this additional
requirement (not due to the fermionic statistics) will be very
important in the derivation of the low-lying spectrum.

Therefore, as a result of the transformations made, our
Hamiltonian has a free oscillator and Coulomb interaction part
only. The interaction part in the old variables with the
introduced cutoff is
\begin{equation}
V \equiv V_{A} + V_{E} = \frac{1}{2} \sum_{\sigma, |\vec{q}| < Q}
\rho_{\sigma}^{old}(-\vec{q}) V_{A}(\vec{q}) \rho_{\sigma}^{old}(\vec{q}) +
                         \frac{1}{2} \sum_{\sigma, |\vec{q}| < Q}
\rho_{\sigma}^{old}(-\vec{q}) V_{E}(\vec{q})
\rho_{-\sigma}^{old}(\vec{q}), \label{vae}
\end{equation}
with $V_{A}(\vec{q}) = 2 \pi e^{2}/|\vec{q}| $ and $V_{E}(\vec{q})
= (2 \pi e^{2}/|\vec{q}|) \exp\{- d |\vec{q}|\}$, where $d$ is the
distance between layers. If we introduce
\begin{equation}
V_{c}(\vec{q}) = V_{A}(\vec{q}) + V_{E}(\vec{q}) \; \;
{\rm and}
\; \;
V_{s}(\vec{q}) = V_{A}(\vec{q}) - V_{E}(\vec{q}),
\end{equation}
and
\begin{equation}
\rho_{c}^{old}(\vec{q}) = \rho_{\uparrow}^{old}(\vec{q}) + \rho_{\downarrow}^{old}(\vec{q})\;\;
{\rm and} \;\;
\rho_{s}^{old}(\vec{q}) = \rho_{\uparrow}^{old}(\vec{q}) - \rho_{\downarrow}^{old}(\vec{q}),
\end{equation}
we can rewrite Eq. (\ref{vae}) as
\begin{equation}
V= \sum_{|\vec{q}|< Q} \frac{1}{4} \rho_{c}^{old}(-\vec{q}) V_{c}(\vec{q}) \rho_{c}^{old}(\vec{q}) +
   \sum_{|\vec{q}|< Q} \frac{1}{4} \rho_{s}^{old}(-\vec{q}) V_{s}(\vec{q}) \rho_{s}^{old}(\vec{q}).
\label{vcs}
\end{equation}
To get the expression in new variables for $\rho_{c}^{old}(\vec{q})$ we should compare
 Eq. (\ref{sden}), Eq. (\ref{newd}), and Eq. (\ref{cdef}) and find
\begin{equation}
\rho_{c}^{old}(\vec{q}) = \frac{q}{\sqrt{4 \pi (2 s + 1)}}
[A(\vec{q}) + A^{\dagger}(\vec{q})] - i  \sum_{j,\sigma} (\vec{q}
\times \vec{p}_{j,\sigma}) \exp\{- i \vec{q}\cdot
\vec{r}_{j,\sigma}\} \label{dnex}
\end{equation}
and also, as we already found out, we have
\begin{equation}
\rho_{s}^{old}(\vec{q}) = \rho_{s}(\vec{q}).
\end{equation}
Further decoupling of the oscillators and particles in $V$ would
amount to higher-order corrections to the expressions found
\cite{msh,sha} and we can safely neglect the presence of
oscillators (terms with $A$'s) in Eq. (\ref{dnex}) when discussing
the low-energy excitations. Then the charge part [the first term
in Eq. (\ref{vcs})] can be decomposed into a diagonal and
off-diagonal part. The diagonal part can be rewritten as
\begin{equation}
\sum_{j,\sigma} \frac{|\vec{p}_{j,\sigma}|^{2}}{2 m_{c}},
\end{equation}
i.e., as a kinetic term of particles with mass $m_{c}$, where
\begin{equation}
\frac{1}{m_{c}} =  \sum_{|\vec{q}| < Q} \frac{V_{c}(\vec{q})}{2} q^{2}
\sin^{2}{\theta_{\vec{q},\vec{p}_{j,\sigma}}}
\label{mss}
\end{equation}
Therefore we came to a description of the system in terms of quasiparticles with a mass that is due
to interactions. These are the expected bosonic dipole objects \cite{nre,msh} with interaction among
them described by the off-diagonal part. As overall neutral objects they should make Bose
condensate(s) in the ground state and we proceed by taking the Bogoliubov expansion of the
quasiparticle operators $\Psi_{\sigma}, \sigma = \uparrow, \downarrow$
in the second-quantized language as
\begin{equation}
\Psi_{\uparrow} = \sqrt{\frac{n_{0}}{2}} + \eta_{\uparrow} \; \;
{\rm and} \; \; \Psi_{\downarrow} = \sqrt{\frac{n_{0}}{2}} +
\eta_{\downarrow},
\end{equation}
where operators $\eta_{\sigma}, \sigma = \uparrow, \downarrow$
describe the small fluctuations around the mean field value,
$\sqrt{n_{0}/2}$, where $n_{0}$ is the density of particles in
each condensate.  We also introduce
\begin{equation}
\eta_{c} = \frac{\eta_{\uparrow} + \eta_{\downarrow}}{\sqrt{2}} \; \; {\rm and} \; \;
\eta_{s} = \frac{\eta_{\uparrow} - \eta_{\downarrow}}{\sqrt{2}},
\end{equation}
new fields that, as we will find out soon, are appropriate for the low-energy description of the
system.

In terms of the new variables, the constraint is, effectively,
\begin{equation}
\Psi_{\uparrow}^{\dagger} \Psi_{\uparrow} +
\Psi_{\downarrow}^{\dagger} \Psi_{\downarrow} - n = n_{0} - n +
\sqrt{n_{0}} (\eta_{c}^{\dagger} + \eta_{c}) + \eta_{c}^{\dagger}
\eta_{c} + \eta_{s}^{\dagger} \eta_{s} = 0. \label{ks}
\end{equation}
Please note that the equality here should be understood as the
equality of the Fourier transforms of lhs and rhs for $\vec{q}$
small. It is also important to notice that although the constraint
effectively is $\rho_{c}^{old}(\vec{q}) = 0$ for $ \vec{q} \neq
\vec{0}$ in the low-energy sector and constrains the first term in
Eq. (\ref{vcs})  to vanish, the underlying canonical variables
$\eta_{c}$ and $\eta_{s}$ may assume nonzero values. To find them,
especially $\eta_{s}$ in which we are mostly interested, we do the
following decoupling.
 Due to the smallness of $\eta_{c}$ and $\eta_{s}$ the constraint may be
rewritten as
\begin{equation}
\eta_{c}^{\dagger}(\vec{q}) + \eta_{c}(-\vec{q}) \approx 0
\end{equation}
so that fields may effectively decouple, satisfying the constraint
only approximately. As a result, from the first part in Eq.
(\ref{vcs}), by relaxing the constraint, we get a kinetic term for
$\eta_{s}$. There are no other contributions to the second order
in $\eta_{s}$. From the first part of Eq. (\ref{ks}) and in the
spirit of the Bogoliubov expansion we may conclude that $\eta_{c}$
is the field that couples to the external electromagnetic
potential. In our decoupling ansatz $\eta_{c}$ is only very weakly
coupled.
 This coincides with the physical picture that we have for bosonic dipoles
that (as dipoles) they weakly interact with external field and
therefore as a system are incompressible \cite{msh}.

Applying the Bogoliubov expansion again, and neglecting the
difference between $n_{0}$ and $n$,
\begin{equation}
\rho_{s} \equiv \Psi_{\uparrow}^{\dagger} \Psi_{\uparrow} - \Psi_{\downarrow}^{\dagger} \Psi_{\downarrow}
\approx \sqrt{n} (\eta_{s}^{\dagger} + \eta_{s}),
\label{rhos}
\end{equation}
we are led to the following Hamiltonian for $\eta_{s}$ fields,
\begin{equation}
{\cal H}_{s} = \sum_{\vec{q}} \frac{|\vec{q}|^{2}}{2 m_{c}}
\eta_{s}^{\dagger}(\vec{q}) \eta_{s}(\vec{q}) + \frac{n}{4}
\sum_{\vec{q}} [\eta_{s}(-\vec{q}) + \eta_{s}^{\dagger}(\vec{q})]
V_{s}(q) [\eta_{s}(\vec{q}) + \eta_{s}^{\dagger}(-\vec{q})].
\end{equation}
As before the hard core boson constraint makes the $V_{s}(0) = 2
\pi e^{2} d $ part of the interaction ineffective but leaves us
with ${\cal H}_{s}$ that describes an unstable system. Therefore
we must impose separately the hard core constraint of composite
bosons on fields $\eta_{s}$. That amounts locally to the following
requirement,
\begin{equation}
\rho_{s}^{2}(r) = \Psi_{\uparrow}^{\dagger} \Psi_{\uparrow} + \Psi_{\downarrow}^{\dagger}
 \Psi_{\downarrow},
\end{equation}
where we used the hard boson properties $ \Psi^{\dagger}_{\sigma} \Psi_{\sigma}
\Psi^{\dagger}_{\sigma} \Psi_{\sigma} = \Psi^{\dagger}_{\sigma} \Psi_{\sigma}, \sigma =
\uparrow $ and $\downarrow$, and $ \Psi_{\sigma}^{\dagger} \Psi_{\sigma} \Psi_{-\sigma}^{\dagger}
\Psi_{-\sigma} = 0 $. Using the Bogoliubov expansion, Eq. (\ref{rhos}), again this becomes
\begin{equation}
n [\eta_{s}^{\dagger}(r) + \eta_{s}(r)]^{2} = n
\end{equation}
that has to be imposed on $\eta_{s}$ fields. Note that here we
also used the incompressibility property of the system in the
low-energy region for the charge degrees of freedom, on the rhs
irrespective of the length scale [Eq. (\ref{ks}) Fourier
transformed for any $\vec{q}$]. We had to make this assumption
because we are incorporating a piece of short-range physics into
the long-wavelength description. Please also note that this is an
operator identity, where the automatic neglect of the quadratic
terms on the lhs of the equation, in the Bogoliubov expansion, is
not allowed.

The constraint we handle in the usual way, switching to the
Lagrangian formulation with fields $ \eta_{s}, \eta_{s}^{\dagger}
$ and a field $\lambda$ that enforces the constraint \cite{nick}.
The generating functional is
\begin{equation}
{\cal Z} = \int {\cal D} \eta_{s} \int {\cal D} \eta_{s}^{\dagger}
\int {\cal D} \lambda \; \exp\left\{ - \int_{0}^{\beta} d \tau
\int d^{2}x \; (\eta_{s}^{\dagger} \partial_{\tau} \eta_{s} +
{\cal H}_{s}(x) + \{[\eta_{s}(x) + \eta_{s}^{\dagger}(x)]^{2} -
1\}\; i \lambda(x,\tau))\right\}, \label{bfir}
\end{equation}
where
\begin{equation}
{\cal H}_{s}(\vec{x},\tau) = \frac{1}{2 m_{c}} \vec{\nabla}
\eta_{s}^{\dagger}\cdot \vec{\nabla} \eta_{s} + \frac{n}{4} \int
d^{2} \vec{y} [\eta_{s}(x) + \eta_{s}^{\dagger}(x)] V_{s}(\vec{x}
- \vec{y}) [\eta_{s}(y) + \eta_{s}^{\dagger}(y)]. \label{bsec}
\end{equation}
The constraint approximately commutes with the Hamiltonian in the
long-wavelength limit (using this property we combined
contributions into a single exponential), and so we will take
$\lambda$, $\tau$ (imaginary time) to be independent. Also, at the
mean-field level, we are allowed to assume that $\lambda$ is space
independent.

Introducing Bogoliubov transformations on $\eta_{s}(\vec{q},\tau)$ fields,
\begin{equation}
\eta_{s}(\vec{q},\tau) = \alpha(\vec{q}) \; \exp\{ i \omega_{q}
\tau \} \; \cosh \theta_{\vec{q}} + \alpha^{\dagger}(-\vec{q}) \;
\exp\{ - i \omega_{q} \tau \} \; \sinh \theta_{\vec{q}}
\end{equation}
[where $\alpha(\vec{q})$ and $\alpha^{\dagger}(\vec{q})$ are new
canonical fields], we get after standard transformations that
diagonalize the problem (see also Ref. \cite{no}),
 the following mean-field expression for $ {\cal Z}$:
\begin{equation}
{\cal Z}_{mf} = \int d\lambda \prod_{\vec{q}}
\frac{1}{1 - \exp\{-\beta \epsilon(\vec{q},\lambda)\}}
\exp\{ - \beta E_{o}(\lambda)\},
\end{equation}
where
\begin{equation}
\epsilon(\vec{q},\lambda) = \sqrt{(\epsilon^{c}_{q})^{2} + [n
V_{s}(q) + 4 \; i \; \lambda] \epsilon_{q}^{c}},
\end{equation}
 with
$\epsilon_{q}^{c} = q^{2}/2 m_{c},$
 and  the domain of $\vec{q}$'s is again the disk with radius $Q$. Also
\begin{equation}
E_{0}(\lambda) = \frac{1}{2} \sum_{q} [ \epsilon(\vec{q},\lambda)
- \epsilon_{\vec{q}}^{c}] - i \; \lambda \sum_{q}
\end{equation}
with the $q$ summations where the cutoff $Q$ is understood. The
$\epsilon(\vec{q},\lambda)$'s are the usual Bogoliubov energies,
the results of the Bogoliubov transformation, now requiring also a
suitable $\lambda$ to get the final expression for the mode
dispersion we are looking for. We approximate $\lambda$ in the
saddle point approach (see, for example, Ref. \cite{assa})
searching for a stationary point of $F(T,\lambda)$, from the
following expression for ${\cal Z}_{mf}$,
\begin{equation}
{\cal Z}_{mf} = \int d\lambda \exp\{-\beta F(T,\lambda)\},
\end{equation}
i.e., look for the solution of
\begin{equation}
\frac{\partial F(T,\lambda)}{\partial \lambda} = 0,
\end{equation}
which effectively becomes
\begin{equation}
\frac{\partial E_{0}(\lambda)}{\partial \lambda} = 0, \label{eq}
\end{equation}
in  the $T \rightarrow 0$ limit.

We solved the equation numerically finding only solutions
$\lambda_{0}$ with $ i \; \lambda_{0}$ real and positive
(therefore, as usual \cite{assa}, we found a path and a saddle
point in the complex $\lambda$ plane), and results are depicted in
Fig. 1.
\begin{figure}
\includegraphics{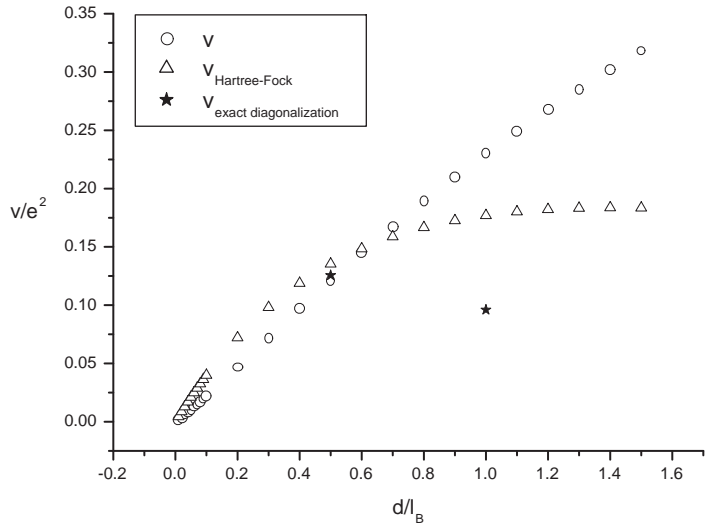}
\caption{The stationary point values for the velocity of the
out-of-phase mode ($\circ$), the values from the pseudospin theory
in the Hartree-Fock approximation from Ref. \cite{moo}
($\triangle$), and exact diagonalization results ($\star$) from
Ref. \cite{moo}.
} \label{one}
\end{figure}
In Fig. 1 we also plotted the Hartree-Fock result of Ref.
\cite{moo}, and the same reference exact diagonalization results at $ d = 0.5
l_{B} $ and $ d = 1l_{B} $. To a good approximation we can claim a
linear dependence for small $d$ of the Bogoliubov velocity though
with the values significantly reduced from the Hartree-Fock
results. But at $d = 0.5 l_{B}$ all three data points are very
close to each other. For larger values of $d$, $d \sim 1 l_{B}$
and larger, both approximation schemes fail to capture the quantum
fluctuations that increase with $d$ \cite{moo}.

The result for $\epsilon(q)$, the dispersion of the out-of-phase
mode in the $d = 0$ case,
\begin{equation}
\epsilon(q) = \epsilon_{q}^{c} = \frac{q^{2}}{2 m_{c}},
\end{equation}
is fairly close to the estimate of Refs. \cite{moo} and
\cite{son}.
Namely $\epsilon(q) =  e^{2} l_{B} \frac{1}{6} \sqrt{2} q^{2} $
while the Hartree-Fock result is $e^{2} l_{B} (\sqrt{\pi}/8)
\sqrt{2} q^{2}$. It is interesting to note that if we use the
expression for $m_{c}$ conjectured in the generalized theory
\cite{sha,rev} that includes higher-momentum physics, we exactly
get the Hartree-Fock result.

It is also interesting to speculate about the discrepancy between
our and the Hartree-Fock result for small $d$. Part of it might be
due to our low-energy, low-momentum limited approach, but it might
also well be due to the incompletness of the
 underlying  analogy \cite{fer,gir} of the $d \neq 0$ system
description compared to the one of a repulsively interacting Bose
gas. (The analogy of the $d = 0$ case to an ideal Bose gas is
complete, as we found out.) The incompletness might follow from
the modifications of the composite boson picture due to the
presence of composite fermions as proposed in Ref. \cite{shs}.
The composite fermions come into relevance very soon as $d$
acquires a nonzero value, and their number rapidly increases with
$d$ \cite{shs}.
If we are allowed to view bosons and fermions to a first
approximation as weakly interacting through a short-range
interaction (more precisely here interacting are differences
between up and down bosons and fermions, respectively) we can
borrow considerations applied to the Bose-Fermi mixtures in
optical traps; see, for example, Refs. \cite{vg}
and \cite{yip}.
When we have a small number of fermions the Bogoliubov mode can,
in fact, enhance its value (on the other hand, decrease) and
damping follows when fermions proliferate. In that sense we can
expect that by considering fluctuations around our mean field
solution (which maps the problem to an interacting bose gas with a
Bogoliubov mode), the value of the Bogoliubov velocity may
increase for small $d$, stay almost the same for intermediate
values, and decrease and even acquire damping for large $d$. In
this sense here we have set up a formalism necessary to check
these considerations in a future work.

It is important to address the case when we do not have the same
density of particles in the up and down layer \cite{the,exp}. Then
in general we have, instead of Eq. (\ref{newsd}),
\begin{equation}
\rho_{s}^{old} = \rho_{s} - (\nu_{\uparrow} - \nu_{\downarrow})
\frac{i}{2} \sum_{j,\sigma} (\vec{q} \times \vec{p}_{j,\sigma})
\exp\{- i \vec{q}\cdot \vec{r}_{j,\sigma}\}, \label{fir}
\end{equation}
when, as before, we neglected the magnetoplasmon part. [
$\nu_{\uparrow}$ and $\nu_{\downarrow}$ ($\nu_{\uparrow} +
\nu_{\downarrow} = 1$) are the filling factors of each layer
separately.] For $\rho_{\sigma}, \sigma = \uparrow, \downarrow $
in $\rho_{s} = \rho_{\uparrow} - \rho_{\downarrow}$ we assume the
following form:
\begin{equation}
\rho_{\sigma}= - \frac{i}{2} \sum_{j} (\vec{q} \times
\vec{p}_{j,\sigma}) \exp\{- i \vec{q}\cdot \vec{r}_{j,\sigma}\} +
C \sum_{j} (\vec{q} \cdot \vec{p}_{j,\sigma}) \exp\{- i
\vec{q}\cdot \vec{r}_{j,\sigma}\}. \label{ass}
\end{equation}
The second part is the longitudinal component of the paramagnetic
current, and the term should appear, in general, when compressible
low-lying degrees of freedom are present. If $C$ in Eq.
(\ref{ass}) is the same for both layers then $ \rho^{c} =
\rho_{\uparrow} + \rho_{\downarrow}$, because the total component
of the current is zero in the charge channel due to its
incompressible nature. Substituting Eq. (\ref{fir}) with Eq.
(\ref{ass}) in the projected Hamiltonian [Eq. (\ref{vcs})] and
collecting all diagonal terms of the dipole expansion, we get
\begin{equation}
\frac{1}{m_{\delta \nu}} = \sum_{|\vec{q}| < Q} q^{2}
\sin^{2}(\theta_{\vec{q},\vec{p}_{j,\sigma}}) \; \left\{
\frac{V_{c}}{2} -  \frac{V_{s}}{2}
(\frac{\nu_{\uparrow}-\nu_{\downarrow}}{2})^{2}\right\} \;
\frac{1}{4} \; \left(\frac{1}{\nu_{\uparrow}} +
\frac{1}{\nu_{\downarrow}}\right), \label{genmass}
\end{equation}
as a generalization of Eq. (\ref{mss}) to the case of imbalanced
layers. This is the mass of the $\eta_{s}$ field defined as $
\eta_{s} = \sqrt{\nu_{\uparrow}} \eta_{\uparrow} -
\sqrt{\nu_{\downarrow}} \eta_{\downarrow}$. So we assumed that we
can apply the Bogoliubov theory and with neglect of some residual
dipole-dipole interaction in the pseudospin channel, our problem
reduces to the one expressed in Eq. (\ref{bfir}) and
Eq. (\ref{bsec}) in which instead of the mass, $m_{c}$, we have
$m_{\delta \nu}$. The assumption is based on the expectation that
the pseudospin channel is compressible. For not large
$|\nu_{\uparrow} - \nu_{\downarrow}| \equiv \delta \nu $, the
velocity of the Bogoliubov mode decreases quadratically with
$\delta \nu$ as a consequence of Eq. (\ref{genmass}), in agreement
with Ref. \cite{jogmac}.
A more detailed investigation of the
influence of the dipole-dipole interaction is needed for general
$\delta \nu$.

We would like to address also the case of huge imbalance, when we
take, for example, $\nu_{\uparrow} \gg \nu_{\downarrow}$
\cite{ky}. As $\nu_{\downarrow} \rightarrow 0, \; 1 / m_{\delta
\nu} \rightarrow \infty,$ and $ \eta_{s} = \sqrt{\nu_{\uparrow}}
\eta_{\uparrow} - \sqrt{\nu_{\downarrow}} \eta_{\downarrow}
\approx \eta_{\uparrow}$, which, probably signals the
incompressible physics of the $\uparrow$ layer. Therefore, to find
out more about the physics of $\downarrow$ quasiparticles, we must
go back to the beginning formulation, and apply a different
decomposition. Namely we will take ( in the limit $\nu_{\uparrow}
\gg \nu_{\downarrow}$)
\begin{equation}
\rho_{s}^{old} \approx \rho_{c}^{old} - 2 \rho_{\downarrow} = - i
\sum_{j,\sigma} (\vec{q} \times \vec{p}_{j,\sigma}) \exp\{- i
\vec{q}\cdot \vec{r}_{j,\sigma}\} - 2 \rho_{\downarrow}.
\label{eks}
\end{equation}
It is easy to see that the first term in Eq. (\ref{eks}) would
lead to the effective mass for all quasiparticles, in the first
approximation, equivalent to what we would have if there was only
one single layer with $\nu = 1$. Next considering the cross term,
\begin{equation}
 \sum_{q < Q} 2 [ - 2 \rho_{\downarrow}(-q)] \frac{V_{s}}{4} \rho_{c}^{old}(q),
\label{cro}
\end{equation}
and taking the expression in Eq. (\ref{ass}) for
$\rho_{\downarrow}$, we get from Eq. (\ref{cro}) for the
$\downarrow$ quasiparticle mass,
\begin{equation}
\frac{1}{m_{e}} =  \sum_{q < Q} V_{E}(\vec{q}) q^{2}
\sin^{2}{\theta_{\vec{q},\vec{p}_{j,\sigma}}}.
\label{masa}
\end{equation}
Because of the incompressible $\uparrow$ background we can neglect
$\uparrow$ and $\downarrow$ cross terms. If we again also assume
irrelevance of the remaining $\downarrow$ dipole-dipole
interactions (for the low-momentum physics), our effective
Hamiltonian for $\downarrow$ quasiparticles is
\begin{equation}
{\cal H}_{e} = \sum_{q < Q} \frac{1}{2 m_{e}} \Psi_{\downarrow}^{\dagger} \vec{p}^{2} \Psi_{\downarrow}
+  \sum_{q < Q} \rho_{\downarrow} \frac{2 V_{s}(q)}{2} \rho_{\downarrow}.
\label{eham}
\end{equation}
In Ref.  \cite{ky}
a physical picture of a Bose gas of excitons and dipoles with
density $n_{\downarrow}$ was developed for the case
$\nu_{\uparrow} \gg \nu_{\downarrow}$. The mass $m_{e}$ we derived
is the result of the low-momentum theory. As in the $ d = 0$,
$n_{\uparrow} = n_{\downarrow}$ case we expect that in the
generalized theory \cite{sha,rev} the cutoff in Eq. (\ref{masa})
would be replaced by a Gaussian in the momentum space and the
expression would coincide with the one in Ref. \cite{ky}.
In this sense with the assumption made, also in this case due to
the comparison to Ref. \cite{ky},
we can claim a complete analogy to a weakly interacting Bose gas
in $ d \rightarrow 0$ limit.


References \cite{fer,mac,moo} and Ref.
\cite{jog}
(when not considering spiral states) agree on the dependence of
the Bogoliubov velocity (in the $n_{\uparrow} = n_{\downarrow}$
case). Possible additions of quantum fluctuations to this value
can be extracted from Ref. \cite{burone}. There, due to the justified
assumption of the suppression of charge fluctuations; a Schwinger
boson mean field theory was used with the requirement on the
single occupancy of the Schwinger boson in a lowest-Landau-level
basis. In this work we were primarily concerned with the
establishment of the concept of a composite boson, and we only set
up the stage for considering fluctuations beyond generalized CS
mean-field theory that is based on this concept. (A composite
boson approach may prove useful for the study of quantum phase
transitions in the bilayer \cite{yetr,shs}, and building of the
physical picture of the bilayer in analogy with the picture based
on composite fermions in the single layer.) Our mean-field theory
and the usual theory do not agree somewhat, although they agree to
a much better degree than the usual CS theory \cite{ezi,ye}
(linear dependence on small $d$ and the absence of the bare mass).
Inclusion of the fluctuations in our hard-core (belonging to
different layer) CS boson model that is probably related to the
model with the single occupancy of Schwinger bosons and comparison
to Ref. \cite{burone}
are planned for future work. It would be important to probe the
significance of the fluctuations around
 $d \sim l_{B}$. Any strong instability
 of the Bogoliubov mode velocity would signal, in the composite-boson-
composite-fermion model (see above and Ref. \cite{shs}),
 the phase separation of the two fluids \cite{stha} and the proposed
first-order transition \cite{schl,burtwo}. Then, from the
composite-boson point of view, we would be able to address in more
detail the extraodinary experiments done on the bilayer
\cite{sp,kg}.
\\


The authors thank N. Read and S.H. Simon for very useful conversations. M.V.M. thanks
Aspen Center for Physics for hospitality during the time when this work has been initiated
and Lucent Bell Labs for their hospitality. The work is supported by Grant No. 1899 of
the Serbian Ministry of Science.

\end{document}